\documentclass[twocolumn,times]{aastex631}
\usepackage{graphicx} 
\begin{document}
\title{Broad band spectral modeling of M87 nucleus }

\correspondingauthor{Andrzej Niedźwiecki}
\email{andrzej.m.niedzwiecki@gmail.com}

\author[0000-0002-8541-8849]{Andrzej Niedźwiecki}
\affiliation{Faculty of Physics and Applied Informatics, Łódź University, Pomorska 149/153, 90-236 Łódź, Poland}

\author[0000-0001-7606-5925]{Michał Szanecki}
\affiliation{Faculty of Physics and Applied Informatics, Łódź University, Pomorska 149/153, 90-236 Łódź, Poland}

\author[0000-0002-1622-3036]{Agnieszka Janiuk}
\affiliation{Center for Theoretical Physics, Polish Acadecmy of Sciences, Al. Lotnikow 32/46, 02-668 Warsaw, Poland}
%\maketitle

\begin{abstract}
We study spectra produced by weakly accreting black hole (BH) systems using the semi-analytic advection dominated accretion flow (ADAF) model and the general-relativistic magentohydrodynamic (GRMHD) simulation. We find significant differences between these two approaches related to a wider spread of the flow parameters as well as a much steeper radial distribution of the magnetic field in the latter. We apply these spectral models to the broad-band spectral energy distribution (SED) of the nucleus of M87 galaxy. The standard (in particular, one-dimensional) formulation of the ADAF model does not allow to explain it; previous claims that this model reproduces the observed SED suffer from an inaccurate treatment of the Compton process. The spectra based on GRMHD simulation are in a much better agreement with the observed data. In our GRMHD model, in which we assumed the BH spin $a=0.9$, bulk of radiation observed between the millimeter and the X-ray range is produced in the disk area within 4 gravitational radii from the BH. In this solution, the synchrotron component easily reproduces the spectral data between the millimeter and the UV range, and the Compton component does not violate the X-rays constraints, for $\dot M \la 0.01 M_{\odot}$/year and a relatively strong magnetic field, with the plasma $\beta \sim 1$ in the region where radiation is produced. However, the Compton component cannot explain the observed X-ray spectrum. Instead, the X-ray spectrum can be reproduced by a high energy tail of the synchrotron spectrum if electrons have a hybrid energy distribution with a $\sim 5$\% non-thermal component. 
\end{abstract}


\section{Introduction}

Accretion flows powering the black hole (BH) X-ray binaries and active galactic nuclei (AGNs) occur in two main forms. At high luminosities, $\ga 1\%$ of the Eddington luminosity, they form optically thick disks, radiating away $\sim 10\%$ of the rest-mass energy of accreting matter. At lower luminosities, the characteristic spectral features of an optically thick material disappear and the radiative efficiency drops significantly. The prevailing interpretation of these low-luminosity BH systems invokes a tenuous, two-temperature mode of accretion \citep{1976ApJ...204..187S,1995ApJ...452..710N,1996ApJ...471..762A}, commonly referred to as an advection-dominated accretion flow (ADAF). The theory of these optically thin flows predicts that a central black hole should be surrounded by a hot, magnetized plasma emitting considerable amounts of synchrotron radiation. These predictions are markedly consistent with the Event Horizon Telescope (EHT) images of  M87* and Sgr A* \citep{2019ApJ...875L...1E,2022ApJ...930L..12E}, in which the observed brightness 
at the EHT frequency, 230 GHz, can be naturally explained by synchrotron emission.
Also, the morphology seen in these images indicates that the near-horizon region is directly observed without obscuration, thus confirming the optical thinness of the radiating material.
Despite these theoretical and observational advances, we still rather poorly understand the details of the physical mechanisms involved, or even whether the observed radiation is produced in the accretion disk or in the inner jet. On the theoretical side, a major source of uncertainty in the quantitative predictions of the expected radiation is the lack of understanding of the physical processes involved in the transfer of the dissipated energy to the plasma particles, and hence how the plasma properties determine the heating efficiency of the radiating particles and how many of them are accelerated into a non-thermal population.

Much work has gone into developing semi-analytical ADAF models \citep[e.g.][]{1997ApJ...489..791M,2000ApJ...534..734M,2012MNRAS.427.1580X,2012MNRAS.420.1195N,2014MNRAS.438.2804N}, following the original formulation by \citet{1976ApJ...204..187S} and \citet{1977ApJ...214..840I}. These models typically involve a detailed treatment of the energy balance of electrons, including the relevant radiative cooling processes as well as advection of energy, compressive heating, and the phenomenological description of the heating efficiency (given by a parameter commonly denoted by $\delta$). The ADAF model has been applied to many low luminosity BH systems, including M87* and Sgr A* which, being one of the best-studied AGNs emitting many orders of magnitude below the Eddington limit,  have been one of the primary targets for the application of this model. 
In this work we revisit these models and we note that some simplifications underlying the analytical descriptions (in particular, the neglect of the multidimensional structure) are a major hindrance to the accurate estimation of spectral formation in the studied objects. 

Magnetohydrodynamic (MHD) simulations of accretion driven by the magnetorotational instability \citep[MRI;][]{1991ApJ...376..214B, 1998RvMP...70....1B} enable a more sophisticated approach to studying flow dynamics and the involved MHD processes. 
Magnetic fields play an important role in transporting the angular momentum outwards, while the accretion flow drags the magnetic field towards the BH horizon. One of the key predictions of general-relativistic (GR) MHD simulations is that the black hole spin must play a key role in powering relativistic jets, while the poloidal magnetic field loops are brought twisted by the frame-dragging effect \citep{2012MNRAS.423.3083M, 2019ARA&A..57..467B}. These jets are directly observed in about 10\% of active galaxies, and one of the most prominent cases is that in M87* galaxy \citep{2023Natur.616..686L}. 
A powerful jet is obtained when the magnetic flux on the horizon
overcomes the gravitational expulsion \citep{2007MNRAS.377L..49K}. However, when the magnetic flux exceeds some threshold value, being a function of the mass accretion rate, the backreaction of the BH magnetosphere acts in the opposite way. The term 
magnetically-arrested disk (MAD) has been introduced by \cite{2011MNRAS.418L..79T} and is frequently used to describe episodic modes of accretion, connected with flaring eruptions of magnetic flux and azimuthally distributed channels of gas entering the black hole  \citep[cf.][]{2022ApJ...935..176J}.
The MAD accretion mode, observed in GRMHD simulations, was in fact predicted already in the calculations based on pseudo-Newtonian framework \citep{2003ApJ...592.1042I, 2008ApJ...677..317I} which showed that
the magnetic fields are dynamically important and change the effectiveness of the MRI. In this mode, the gas is pushed outward by magnetic tension, and the interchange instability is the one which ultimately allows it to accrete through the BH horizon.

While the GRMHD models provide a thorough description of the flow structure, they typically neglect radiative cooling processes, and the treatment of the thermodynamics of the electrons is very cursory. A few radiative GRMHD simulation were carried out with self-consistent evolution of two-temperature plasma \citep[e.g.][]{2018ApJ...864..126R,2019MNRAS.486.2873C} and the results were found to be strongly dependent on the assumptions concerning the electron heating.
In most cases, however, GRMHD simulations evolve a single fluid whose internal energy is dominated by ions, while the temperature of the emitting electrons remains unconstrained and is then typically set manually for radiation calculations.

In this work we compare spectra predicted by the ADAF and the GRMHD models for accretion rates relevant for M87*, using the radiative-transfer code of \citet{2005MNRAS.356..913N} and \citet{2012MNRAS.420.1195N}. Such a comparison has never been made before. We then attempt to reproduce the broadband spectral energy distribution (SED) observed in M87* using these spectral predictions. Specifically, we apply them to the radio to X-ray range data compiled by \citet{2016MNRAS.457.3801P}, and to the results of the quasi-simultaneous multi-wavelength campaign in 2017, reported by \citet{2021ApJ...911L..11E}.

\section{Notation}

We use the usual dimensionless parameters, $r = R/R_{\rm g}$, $a = J / (c R_{\rm g} M)$ and $\dot m = \dot M / \dot M_{\rm Edd}$, where $R_{\rm g}=GM/c^2$ is the gravitational radius, $J$ is the black hole angular momentum, $\dot M_{\rm Edd}= L_{\rm Edd}/(\epsilon c^2)$, $L_{\rm Edd} \equiv 4\pi GM m_{\rm p} c/\sigma_{\rm T}$, and to define the Eddington rate we take the radiative efficiency to be $\epsilon = 0.1$. The BH horizon radius is $r_{\rm hor} = 1 + (1 - a^2)^{1/2}$.

For M87*, we assume the distance of $d = 16.7$ Mpc \citep{2007ApJ...655..144M,2009ApJ...694..556B}, the source inclination, i.e.\ the viewing angle of the emitting region with respect to the line of sight, of
$i = 17^\circ$ \citep[corresponding to the angle between the approaching jet and the line of sight;][]{2018ApJ...855..128W}, and the black-hole mass of $M = 6.5 \times 10^9 M_{\odot}$, as adopted by the EHT Collaboration. This corresponds to $\dot M_{\rm Edd} = 148 \, M_\odot$ year$^{-1}$.                

The plasma $\beta$ parameter is defined as the ratio of the gas to magnetic pressure, $\beta=P_{\rm gas}/P_{\rm mag}$, and the fraction of accretion power used to directly heat electrons is denoted by $\delta$. The number density is denoted by $n$, the mass density by $\rho$, and the electron and ion temperature by $T_{\rm e}$ and $T_{\rm p}$. The Boltzmann's constant is denoted by $k$ and the proton mass by $m_{\rm p}$.

\section{ADAF models for M87*}
\label{sect:adaf}

We first apply a semi-analytic ADAF model {\tt kerrflow} \citep{2022ApJ...931..167N}. It strictly follows the standard ADAF formalism, which is based on transonic solutions of the hydrodynamical equations for the conservation of mass, radial momentum, angular momentum, and separate energy equations for ions and electrons, using the phenomenological $\alpha$-prescription for viscosity \citep[e.g.][]{1997ApJ...476...49N,1997ApJ...489..791M,1998ApJ...498..313G,2000ApJ...534..734M}. The hydrodynamic description underlying these solutions is essentially one-dimensional, giving only the radial distribution of the flow parameters (which turns out to be a major oversimplification, as we find below).
As in all ADAF solutions, the plasma is assumed to be described by radius-independent values of $\alpha$ and $\beta$ parameters. Despite these simplifications, the analytic ADAF models have an important advantage of including a detailed description of the energy balance of the electrons, including the relevant radiative cooling processes (the synchrotron and bremsstrahlung emission and their Comptonization), and are therefore a good starting point for estimating the expected values of the electron temperature. 
These models have been used extensively for the description of black-hole systems and their comparison with models based on MHD simulations can be valuable.

The {\tt kerrflow} model improves on some computational inaccuracies present in other similar models. The most important one concerns the treatment of Comptonization, for which  {\tt kerrflow}  uses an accurate, GR Monte Carlo simulations, whereas in other published models it is computed using the approximate formulae of \cite{1990MNRAS.245..453C}. The \cite{1990MNRAS.245..453C} approximation, however, is only accurate at nonrelativistic temperatures (as clearly stated in that paper) and underestimates the energy increase in a scattering at large $T_{\rm e}$. Then, using it in ADAF models, which are typically characterized by relativistic $T_{\rm e}$, results in inaccurate spectral predictions, to the extent dependent on the model parameters. At $\dot m \sim 10^{-3}$, when the Comptonization spectrum has a power-law shape,   the approximation gives a far too soft spectrum. At $\dot m \la 10^{-4}$, when individual scattering orders are seen in the spectrum, it strongly underestimates the energy at which the scattering bumps are seen. Furthermore, all previous models \citep[with a notable exception of][referred to below]{2009ApJ...699..513L} used a non-GR description (in particular, pseudo-Newtonian potential for hydrodynamics and neglect of GR transfer of radiation), which fails in the region where most of the accretion power is released and radiation is produced. In contrast, {\tt kerrflow} is fully GR. 

\begin{figure*}
\centering
 \includegraphics[width=15cm]{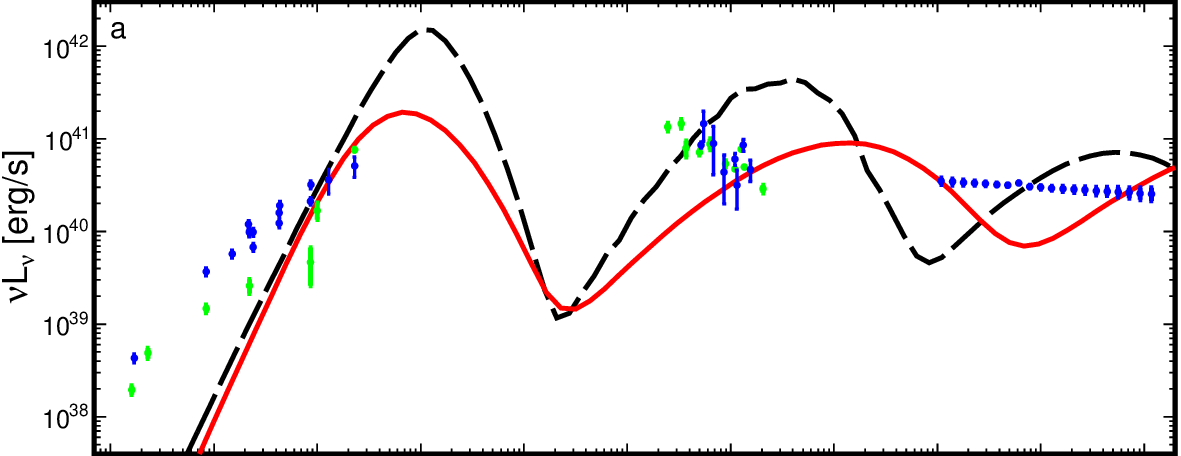}
 \includegraphics[width=15cm]{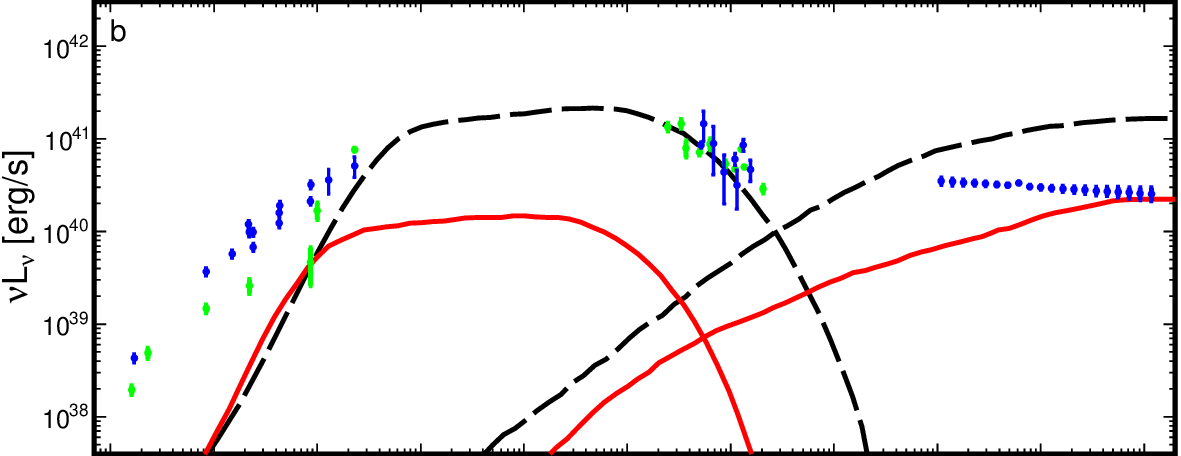}
 \includegraphics[width=15cm]{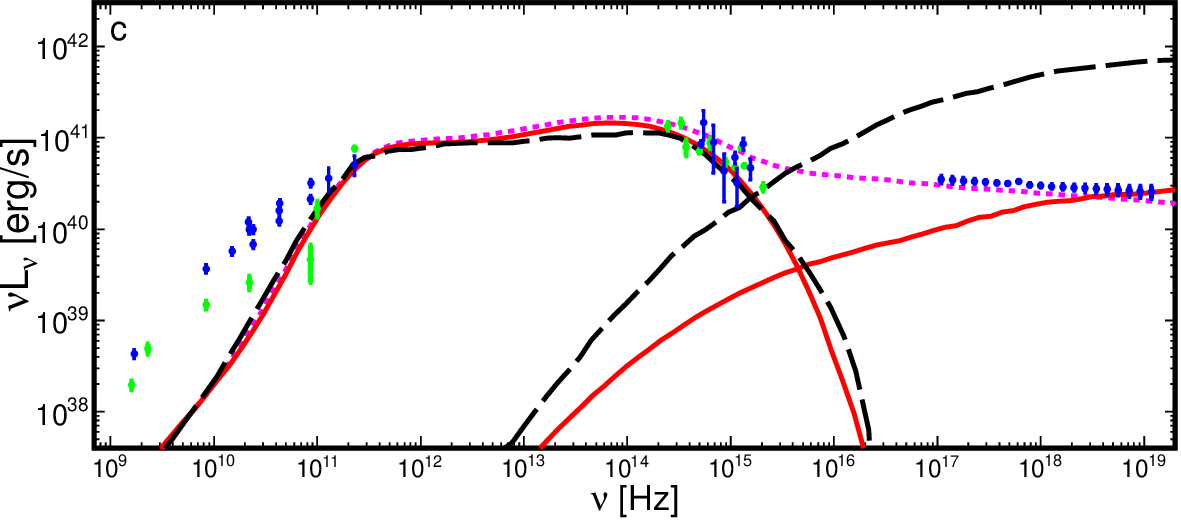}
\caption{Comparison of the observed SED of M87 core with spectra predicted by (a) the analytic ADAF model, and (b,c) the model based on the GRMHD simulation. In all panels the green points show the highest spatial resolution SED in the quiescent phase of M87* given in table 4 of \citet{2016MNRAS.457.3801P}, and the blue points show results of the quasi-simultaneous multi-wavelength campaign in 2017, reported by \citet{2021ApJ...911L..11E}. See Table \ref{tab:models} for definition of the models. All models assume $M = 6.5 \times 10^9 M_{\odot}$ and $i=17^{\circ}$; all spectra except for the dotted magenta line in (c) assume  a purely thermal distribution of electron energy. In panel (a) the solid red line shows the spectrum of the {\tt kerrflow} model A1; the dashed black line shows the spectrum of {\tt riaf-sed} \citep{2014MNRAS.438.2804N} for the same  $\dot m = 10^{-5}$, 
$\delta=0.1$ and $\beta=1$.  The difference between the model spectra is primarily due to the neglect of GR effects and the inaccurate treatment of Comptonization in {\tt riaf-sed}. Panel (b) illustrates how the GR transfer affects the spectrum in the GRMHD model B; the synchrotron and Compton components are shown separately. The dashed black line shows the spectrum in the flat space-time (i.e.\ neglecting the GR transfer) and the solid red line shows the fully-GR spectrum. Panel (c) shows how the Compton component is affected when the synchrotron component is matched to the data by increasing either the magnetic field (the solid red line, model D) or the density and the electron temperature (the dashed black line, model C). The synchrotron and Compton components are shown separately. The dotted magenta line shows the synchrotron spectrum in model D with a hybrid electron energy distribution, including a 5\% energy content of non-thermal electrons in the form of a high-energy power-law tail with an index equal to 3.2.}
\label{fig:sed}
\end{figure*}

The effect of these two shortcomings on the spectrum is illustrated in Figure \ref{fig:sed}(a), where the {\tt kerrflow} spectrum is compared with the one computed for the same parameters with a non-GR model {\tt riaf-sed} \citep{2014MNRAS.438.2804N}. Both models predict very similar values of density and electron temperature (in {\tt riaf-sed}, $T_{\rm e}$ is slightly larger) and then the large difference in the position of the scattering bumps seen in these spectra is entirely due to deficiencies of the \cite{1990MNRAS.245..453C} approach applied in {\tt riaf-sed}. The difference in the shape and amplitude of the synchrotron peak is due to the neglect of both the photon advection by the BH and the GR redshift in 
{\tt riaf-sed}. 
We note, in particular, that all previous attempts to explain the M87* spectrum with ADAF models relied on fitting the first scattering bump to the optical/UV data and the second scattering bump to the X-ray data. The strong underprediction of the energies of these bumps by the applied \cite{1990MNRAS.245..453C} approximation  makes these fitting attempts incorrect.

In its most general version, the ADAF model includes a number of free parameters and exploring the full parameter space is not feasible. We are most interested in estimating the suitable values of $\beta$, $\delta$ and $\dot m$, therefore, we allowed these parameters to vary, while keeping the others fixed. Specifically, we assumed the usual value of $\alpha=0.3$ \citep[as discussed by][there is a degeneracy between $\alpha$, $\dot m$ and $\delta$]{2022ApJ...931..167N}. Furthermore, it has become customary to assume a local accretion rate decreasing inward as a power-law, $\dot m = \dot m_{\rm out} (r/r_{\rm out})^s$,
at all radii down to the black-hole horizon, typically with $s \sim 0.5$. However, assuming such a $\dot m (r)$ profile in the inner flow may strongly affect the spectral properties \citep[cf.][]{2022ApJ...931..167N}, whereas there is no clear indication from MHD simulations that such a strong mass loss should occur in the innermost region (and actually it was not found in the simulations considered in Section \ref{sect:mhd}) as well as  the observational evidence of winds in systems with very low accretion rates concerns the radial scales of $r \sim 10^4 - 10^5$, e.g.\ in M87* \citep{2019ApJ...871..257P}, but not at lower radii. Therefore, we assume that $\dot m$ is constant in the region where the radiation is produced, i.e.\ we set $s=0$. Then, we also assume $a=0.9$, which value may be relevant for a system with a powerful jet.

We found that it is not possible to reproduce the full SED with the {\tt kerrflow} model, therefore, we focused on modeling the 
millimeter part. We considered $1 \le \beta \le 9$ and $0.01 \le \delta \le 0.5$ and we found that the model best-matches the millimeter data for $\dot m \simeq 10^{-5}$ and $\delta \simeq 0.1$, and this result does not depend on the value of $\beta$. These solutions are represented by models A1 and A9 defined in Table \ref{tab:models}, and the solid red line in Figure \ref{fig:sed}(a) shows the spectrum for model A1 with $\beta=1$.
We now compare this result with previous ADAF models for M87*.
The early models for this object \citep{1996MNRAS.283L.111R,2003ApJ...582..133D} used the simplified self-similar description of \citet{1995ApJ...452..710N} and then more accurate ones, based on transonic hydrodynamic solutions, were presented by \citet{2009ApJ...699..513L}, \citet{2014MNRAS.438.2804N} and \citet{2019MNRAS.490.4606B}. Of these, only the model of \citet{2009ApJ...699..513L} is GR and then it is closest to  {\tt kerrflow}. Indeed, we note a very good agreement between parameters and synchrotron spectra of these two models, with the main difference concerning $\sim 3$ times
%\footnote{after scaling to $M$ assumed in this work} 
larger $\dot m$ and correspondingly stronger synchrotron peak in \citet{2009ApJ...699..513L}; this larger $\dot m$ likely results from the (incorrect, see above) attempt to reproduce the data above $10^{14}$ Hz with the Compton component, and the scarcity of radio data available at that time.
\citet{2014MNRAS.438.2804N} and \citet{2019MNRAS.490.4606B} found parameters strongly differing from ours, in particular a large $\dot m \simeq 2 \times 10^{-4}$ in the inner part, which may reflect large computational differences between the models and again their results relied on matching the incorrectly determined Compton spectrum to the data.

\begin{table}[]
    \centering
    %\begin{ruledtabular}
    \begin{tabular}{ c c c c }
    \hline 
         ADAF model & $\dot m$ [$\times 10^{-5}$] & $\beta$ & $\delta$ \\
         \hline 
         A1 & 1 & 1 & 0.1 \\
         A9 & 1 & 9 & 0.1 \\
    \hline 
         GRMHD model & $\dot m$ [$\times 10^{-5}$] & $b_{\rm mag}$ & $T_{\rm e}/T_{\rm p}$ \\
         \hline 
         A & 3.8 & 1 & 0.04 \\
         B & 25 & 0.4 & 0.04 \\
         C & 35 & 0.34 & 0.1 \\
         D & 7.6 & 3.9 & 0.04 \\
    \hline      
    \end{tabular}
    %\end{ruledtabular}
    \caption{Parameters of the models. All models assume $M = 6.5 \times 10^9 M_{\odot}$ and the spin parameter $a=0.9$. 
    }
    \label{tab:models}
\end{table}

\section{Model based on GRMHD simulations}
\label{sect:mhd}

\subsection{Simulations and post-processing}

In this Section we analyze the results of 3D general-relativistic ideal GRMHD
numerical simulations of accretion flows onto a magnetically saturated Kerr BH.
For the analysis, we choose the AGN-HS100  simulation presented in \cite{2022A&A...668A..66J}. Our run was performed with the code \textit{HARM\_COOL} \citep[presented in e.g.][for a 2D case] {2019ApJ...873...12S}.
The MHD equations for the continuity and energy momentum are brought in the code in conservative form in general relativity, and account for the dynamical evolution of magnetized plasma in the gravitational potential of a spinning BH.

Our simulation is performed on a fixed Kerr background metric, with proper coordinate transformation between spherical (Boyer-Lindquist) and code (Modified Kerr-Schield) coordinates, to avoid the flow penetrate the black hole horizon. The inner radius of the grid is located at $r_{\rm in}=1.01 r_{\rm hor}$, and the outer radius is fixed at $10^{5} R_{\rm g}$. The number of points in r-$\theta$-$\phi$ directions id 288x256x256, respectively. Thanks to adopting a progressively sparser grid above $r=400$, our simulations are well resolved close the black hole horizon. 

The initial conditions are imposed by the equilibrium torus solution \cite{FM76}, located initially at $r_{\rm in}= 12$ where the cusp forms, and parameterized with the radius of pressure maximum $r_{\rm max}=25$. BH spin $a=0.9$ is adopted. To start accretion, the initial torus is embedded in a poloidal magnetic field, prescribed by the vector potential of the form $A_{\phi}= r^{5} (\rho/\rho_{\rm max} - 0.2)$, where $\rho$ is the local density in the torus and $\rho_{\rm max}$ is the density maximum. Therefore, the poloidal loops are centered at the radius $r_{\rm max}$. The field is initially set to zero outside the torus. Factor $r^{5}$ enables bringing more magnetic flux from large distances and formation of the MAD state. In addition, the initial distribution of internal energy is perturbed azimuthally, by a random fluctuation, of the form $u=u_{0}(0.95+0.1 {\rm RAND})$ where RAND is a random number uniformly distributed between 0 and 1.

The GRMHD code works in geometrized units and has no intrinsic mass scale and then to apply it to M87* we need to introduce specific scalings
to convert the code variables to cgs units. This is done after adopting the BH mass (i.e.\ $M = 6.5\times10^{9}M_{\odot}$ for M87*) and the resulting scalings for gravitational length and time, $R_{\rm g} = GM/c^2$ and $\mathcal{T} = R_{\rm g}/c$.
The factor scaling the density to the cgs units, $\mathcal{M}$, is then related to the assumed accretion rate by $\mathcal{M} = \dot M \mathcal{T}/R_{\rm g}^3$. 
The accretion rate is measured at the inner boundary, close to the event horizon, and is determined at the final time of simulation when a quasi-stationary state is established.

We have found that it is not possible to reproduce the observed SED by adjusting only the $\dot m$ parameter, and therefore we allow small variations in the scaling of the magnetic field by introducing a free parameter (close to unity), $b_{\rm mag}$, so that the scaling factor for the magnetic field is $\mathcal{B} = b_{\rm mag} (8 \pi \mathcal{M} c^2)^{1/2}$. This approach is motivated by an ambiguity (by a factor of several) in relating the source $\dot M$ to the simulation accretion rate, since the latter shows order-of-magnitude variations during the simulation run. Adjusting $b_{\rm mag}$ to fit the SED then reflects significant variations in the accretion rate at the event horizon, corresponding to much weaker variations in the magnetic field, in the AGN-type simulations on which we rely here, as expressed quantitatively e.g.\ by changes in the MAD-ness parameter (the ratio of the magnetic flux to the square root of the accretion rate) by a factor of several \citep[see Figure 6 in][]{2022A&A...668A..66J}.

Furthermore, the MHD simulation gives only the total internal energy density, $u_{\rm gas}$, and not the distinct
electron and ion temperatures. 
A common practice in radiation calculations in such models, the so-called post-processing, is to set $T_{\rm e}$ after running the
simulation, as a function of local plasma parameters.
As we discuss below,
the popular parameterization used for this, introduced by \citet{2016A&A...586A..38M} and referred to as the $R$-$\beta$ model, 
may yield unrealistically high values of $T_{\rm e}$, as it ignores limitations due to cooling efficiency. Therefore, we do not follow it. 
Instead, we set the electron temperature based on the results obtained in our ADAF model, which includes  a detailed treatment of the electron energy balance and then gives a reasonable indication of the expected magnitude of $T_{\rm e}$ under relevant physical conditions.  

\begin{figure*}
\centering
 \includegraphics[width=7cm]{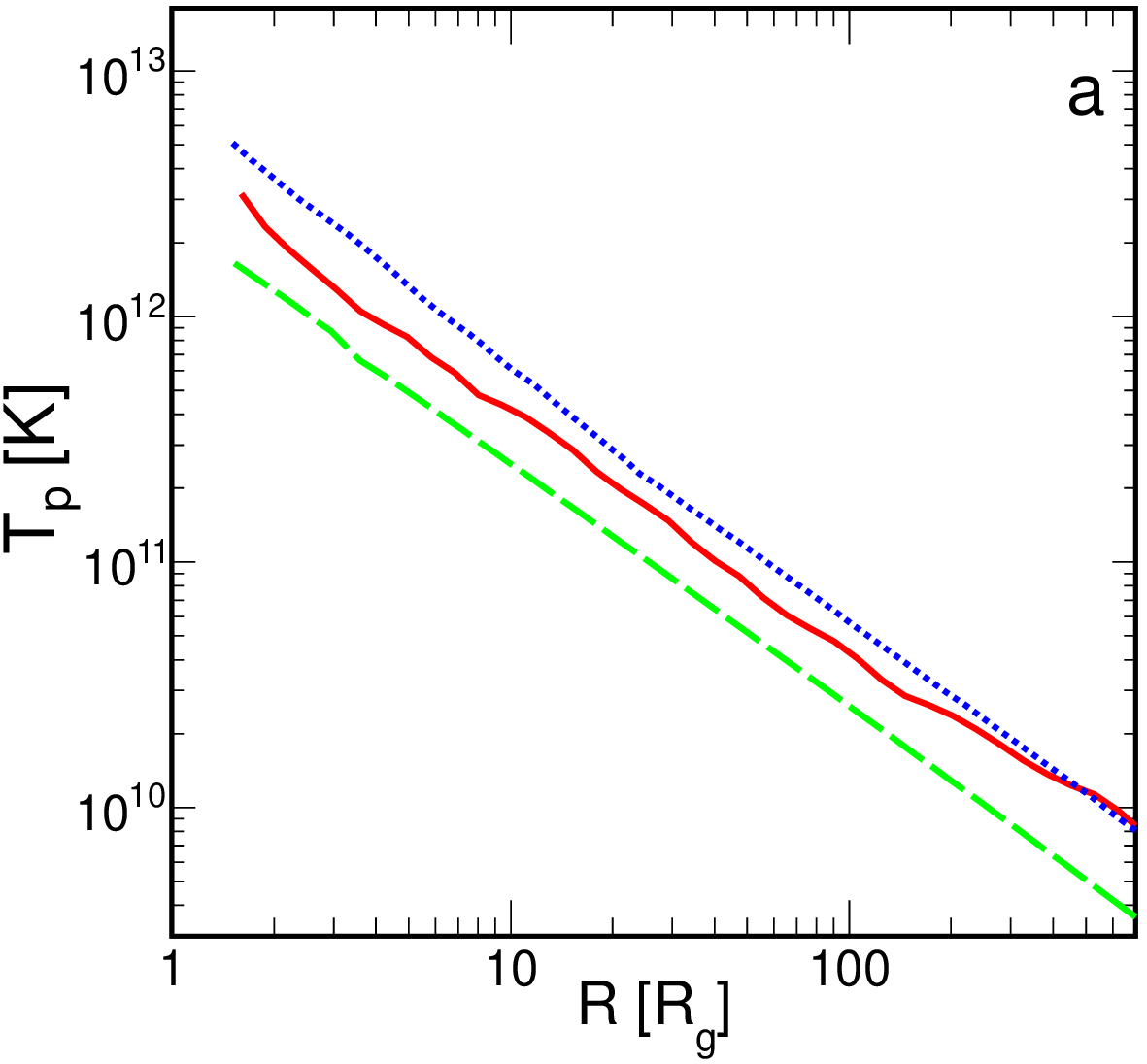}
  \includegraphics[width=7cm]{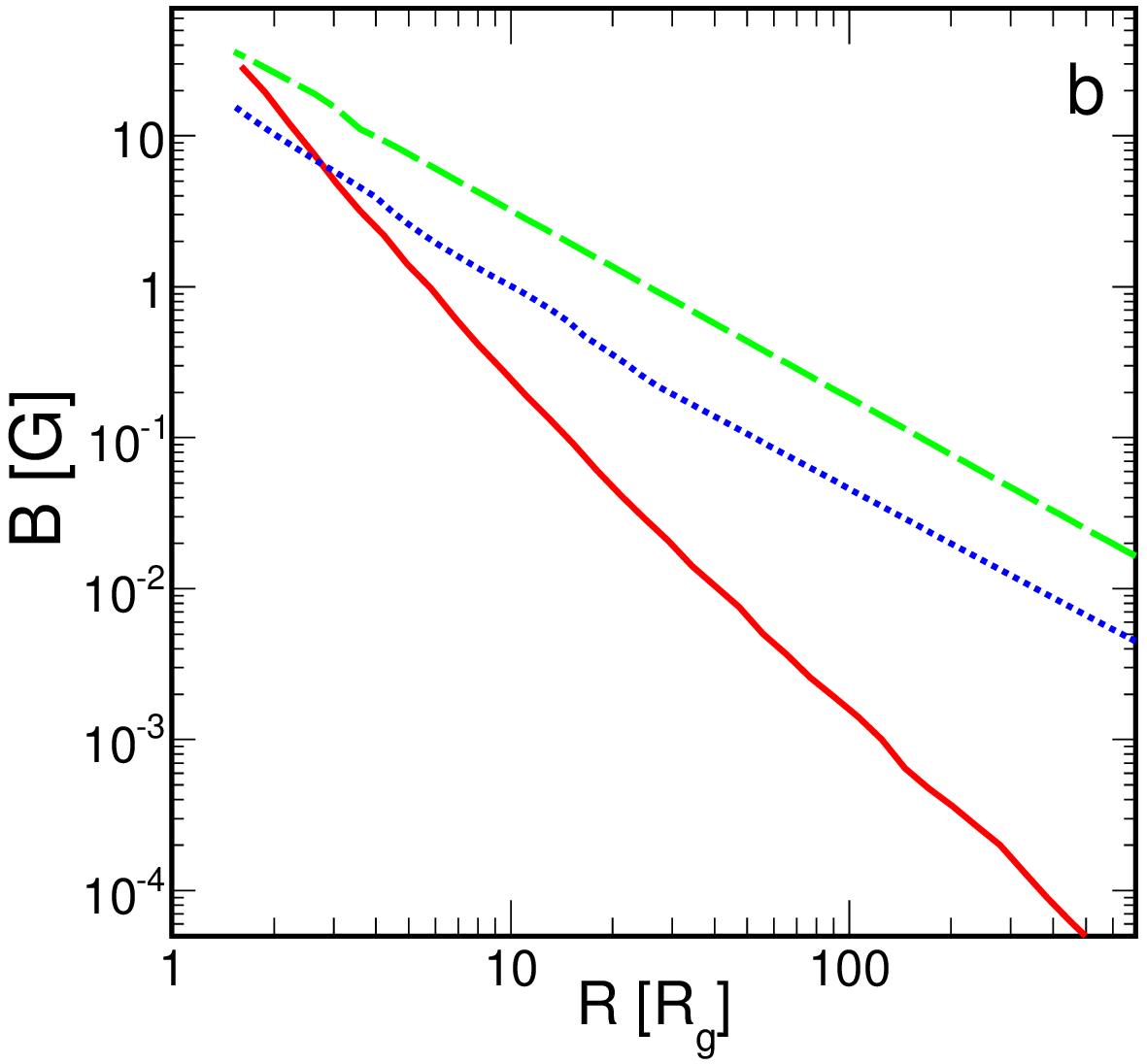}
 \includegraphics[width=7cm]{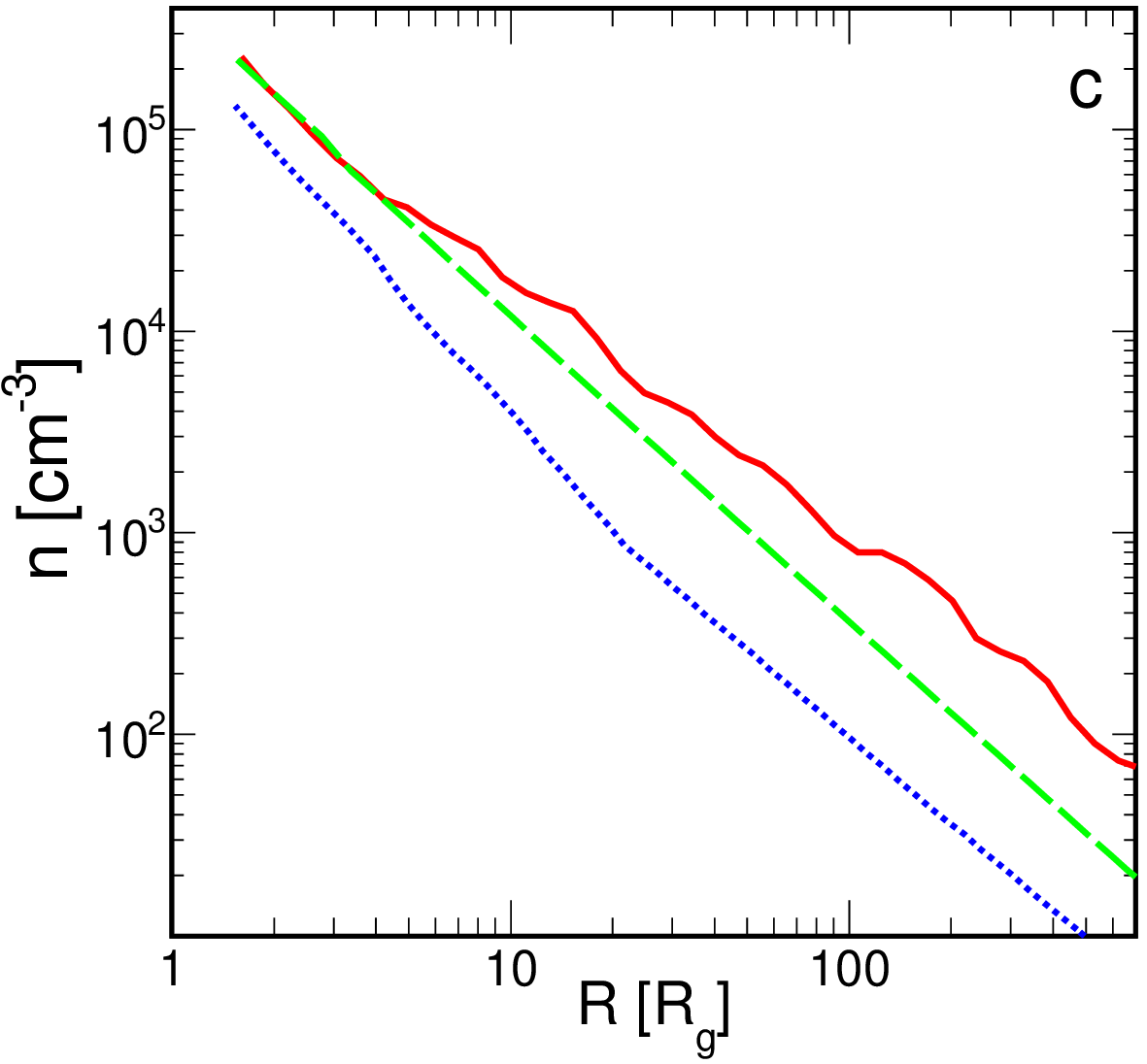}
 \includegraphics[width=7cm]{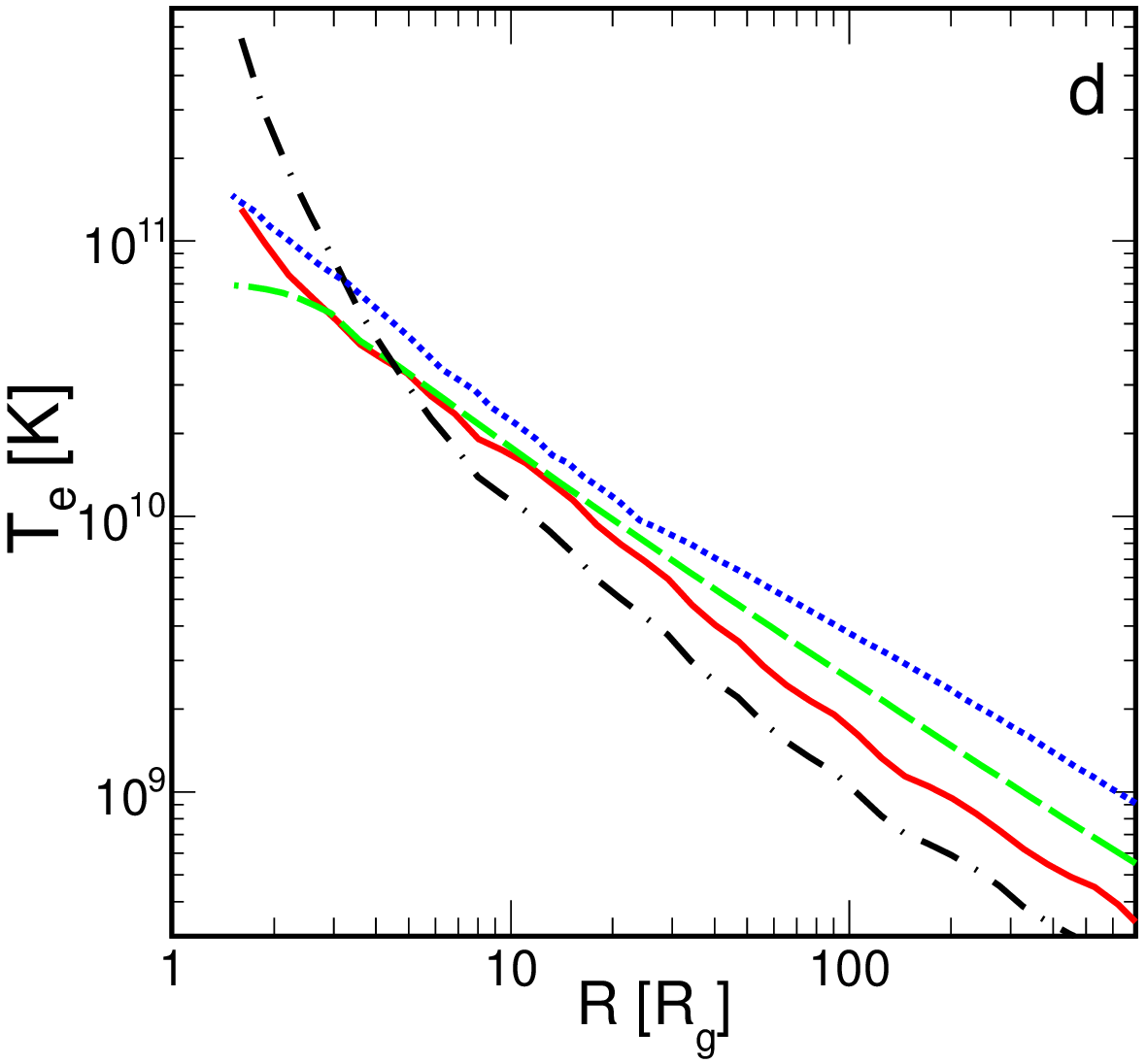}
\caption{Comparison of the radial profiles of (a) $T_{\rm p}$, (b) $B$, (c) $n$, and (d) $T_{\rm e}$ for the {\tt kerrflow} models A1 (the dashed green lines) and A9 (the dotted blue lines) and the GRMHD model A (the red solid lines), see Table \ref{tab:models}. In panel (d), the dot-dashed black line is for the $R-\beta$ relation, equations (\ref{eq:rbeta2}) and (\ref{eq:rbeta1}), with $R_\mathrm{high}=80$, $R_\mathrm{low}=1$; the solid red line is for $T_{\rm e} = 0.04 T_{\rm p}$ assumed in most of our spectral models. 
In the GRMHD model, $T_{\rm p}$, $B$, and $T_{\rm e}$ are found by spherical, density-weighted averaging (over narrow  shells with $\Delta r = 0.1 r$), and
$n$ by averaging over narrow rings (again with $\Delta r = 0.1 r$) with $z \le 0.1 r$.
}
\label{fig:radial}
\end{figure*}

Figure \ref{fig:radial}(a-c) shows the radial profiles of $B$, $T_{\rm p}$ and $n$ in the {\tt kerrflow} models A1 and A9. The lower $\beta$ model A1 is characterized by a higher $B$ (which is a trivial consequence of the assumed higher magnetization), lower $T_{\rm p}$ (because a larger fraction of the accretion power is used to build up the magnetic field instead of heating the flow), and higher $n$ (because a more magnetized flow is geometrically thinner). 
Also shown in Figure \ref{fig:radial}(a-c) are the radial profiles of quantities representing the GRMHD solution, for which the average $\bar{B} (r)$ and $\bar T_{\rm p}(r)$  are found by density-weighted averaging ($\bar B = \int B n {\rm d} V / \int n {\rm d} V$ and analogously for $\bar T_{\rm p}$) over narrow, spherical shells,  and $\bar n(r)$ by averaging ($\bar n = \int n {\rm d} V / \int {\rm d} V$) over narrow rings representing the mid-plane ($z \le 0.1 r$, where $z$ is the distance from the mid-plane). The major difference between the GRMHD simulation and the analytic ADAF calculation concerns $B$, which for the former has a much steeper profile corresponding to a gradual build-up of the magnetic field, in contrast to comparatively flatter profiles resulting from the assumption of constant $\beta$ in the ADAF model. The GRMHD solution also has a slightly flatter profile of $n$, due to the weak loss of accreting matter to the wind, and $T_{\rm p}$ decreasing from a large-$\beta$ to a small-$\beta$ value toward smaller $r$, but with a relative increase close to the event horizon due to the additional heating by rotational energy extraction from the BH (neglected in the ADAF model).
Nevertheless, except for the difference in the gradient of $B$, we see a very good agreement of the parameters characterising the ADAF and MHD solutions. We can therefore expect that both solutions should also be characterised by similar values of $T_{\rm e}$.

Figure \ref{fig:radial}d shows the $T_{\rm e}$ profiles in our models. The temperature ratio in the {\tt kerrflow} solutions A1 and A9 is in the range $T_{\rm e}/T_{\rm p} \simeq (0.02 - 0.07)$  and, more importantly, in the innermost part $T_{\rm e} \simeq 10^{11}$ K (within a factor $\la 2$). 
Guided by these results, in most of our GRMHD radiative models we set $T_{\rm e} = 0.04 T_{\rm p}$ (we checked how the increase of $T_{\rm e}$ affects the spectrum and we found that this leads to a too strong Compton emission, as illustrated in model C defined below).
For the proton temperature
we assume that electrons give a negligible contribution to $u_{\rm gas}$, and then the simulated values of $u_{\rm gas}$ are used to determine $T_{\rm p}$. 
With our adopted prescription for the local $T_{\rm e}$ (i.e.\ $ = 0.04 T_{\rm p}$), the average $\bar T_{\rm e}(r)$ is close to the value corresponding to the energy balance in our ADAF calculations, as shown  in Figure \ref{fig:radial}d.

\begin{figure*}
\centering
 \includegraphics[width=7cm]{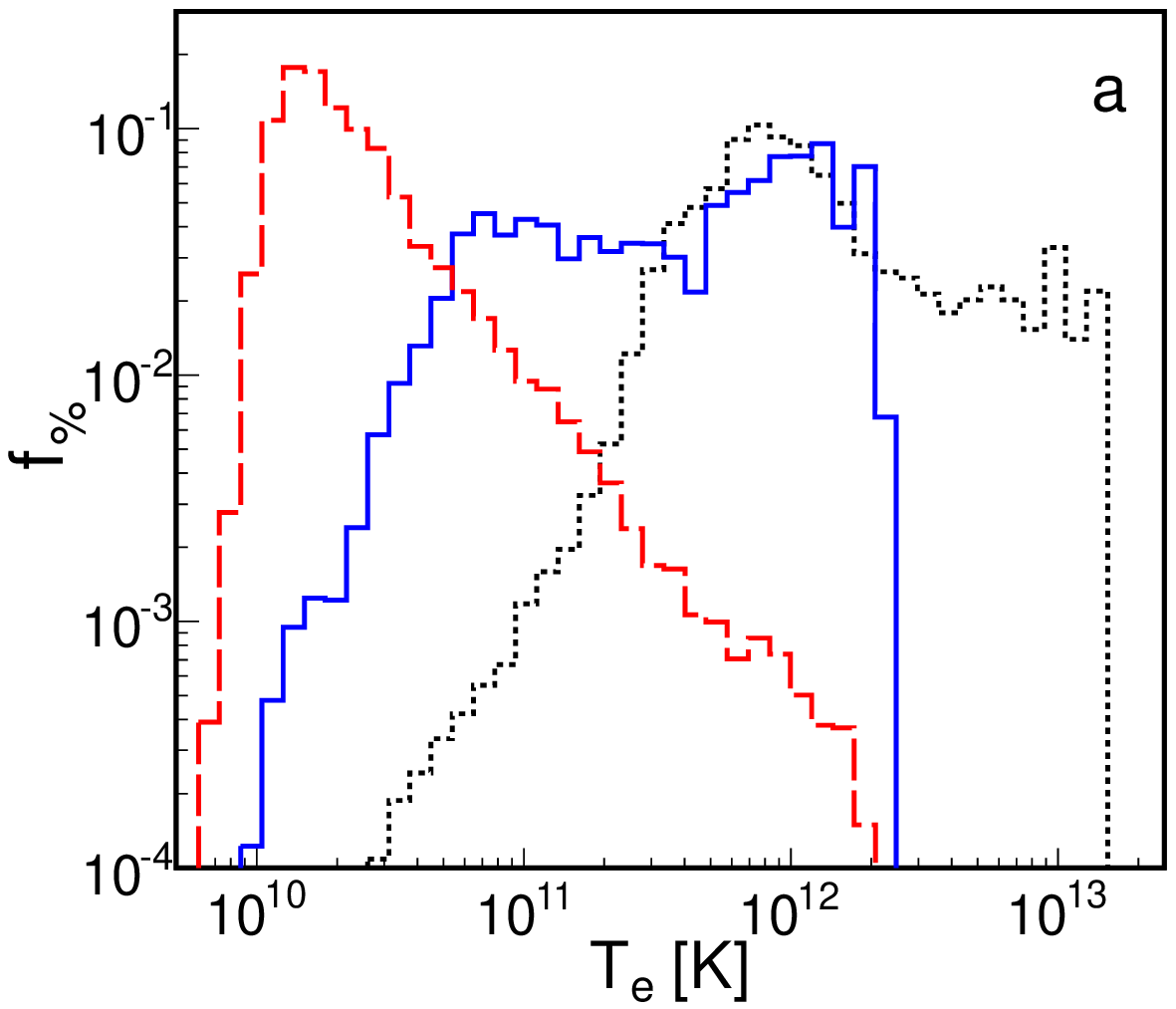}
  \includegraphics[width=7cm]{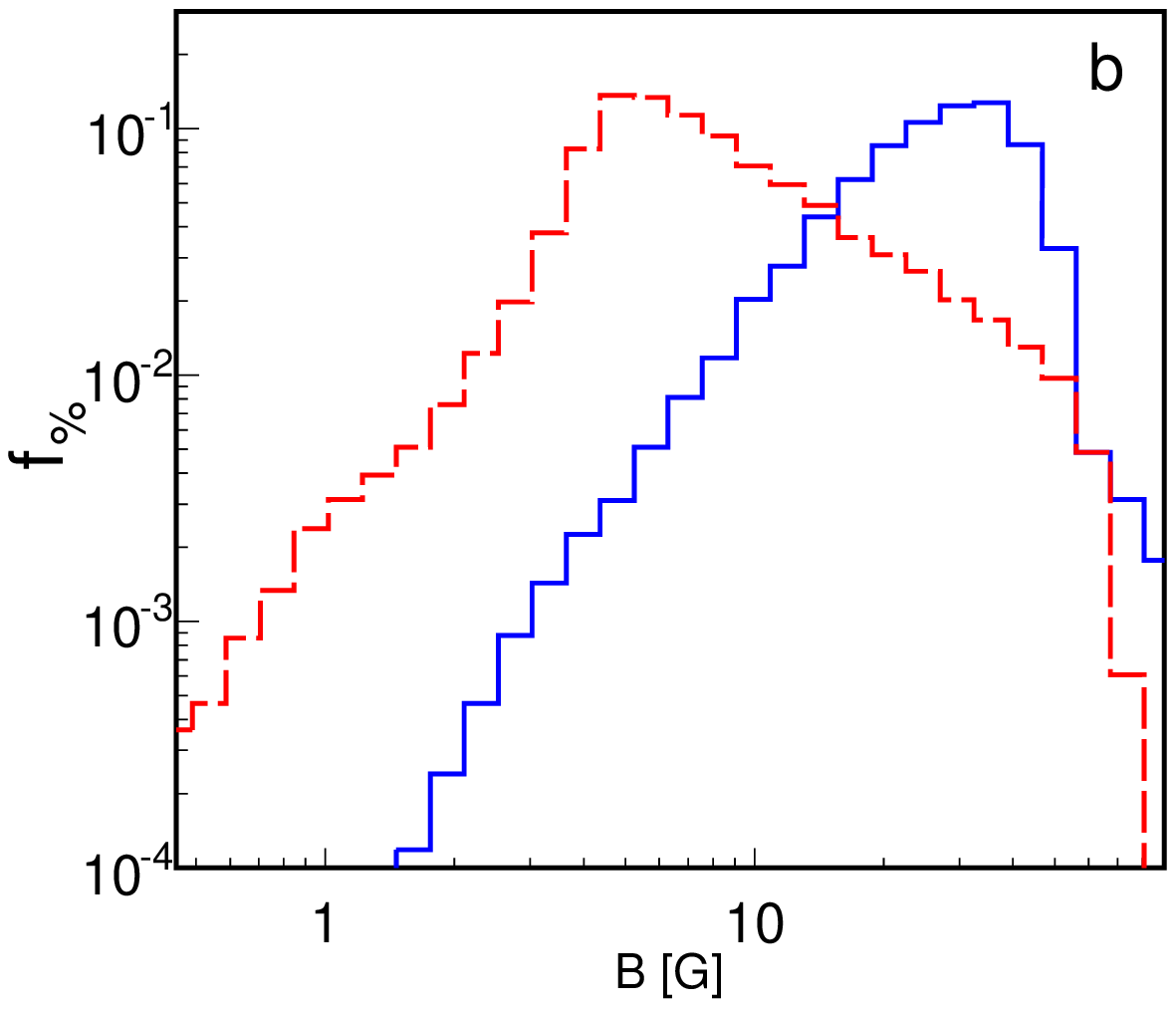}
\caption{The dashed red lines show the volume  filling factors of the regions with a given value of (a) the electron temperature and (b) the magnetic field in model A at $r \le 4$. The solid blue lines show a similar distribution of the regions with a given value of these parameters, but weighted by the synchrotron emissivity, which then gives the contribution of these regions to the total synchrotron emission. The dotted black line in panel (a) shows the distribution of $T_{\rm e}$ weighted by the synchrotron emissivity for the $R-\beta$ relation with $R_\mathrm{high}=80$ and $R_\mathrm{low}=1$.}
\label{fig:new}
\end{figure*}

We now briefly comment on the $R$-$\beta$ model, 
in which the ratio between the ion and electron temperatures is given by the local
plasma $\beta$, 
\begin{equation}
R \equiv \frac{T_{\rm p}}{T_{\rm e}} = R_\mathrm{high} \frac{\beta^2}{1+\beta^2} + R_\mathrm{low} \frac{1}{1+\beta^2},
\label{eq:rbeta2}
\end{equation}
which then allows to relate $T_{\rm e}$ to $u_{\rm gas}$ by
%from the simulation 
\begin{equation}
T_{\rm e} = \frac{2 m_{\rm p} u_{\rm gas}}{3 \rho k (2+R)},
\label{eq:rbeta1}
\end{equation}
see e.g.\ \citet{2019ApJ...875L...5E}.
In  the above $R_\mathrm{high}$ and $R_\mathrm{low}$ are the free parameters of the model and give the temperature ratios at high and low $\beta$, respectively.
The $R$-$\beta$ prescription tentatively relates the temperature distribution to the efficiency of the electron heating, which increases with plasma magnetisation, as often seen in simulations of electron heating in turbulent or reconnecting plasmas \citep[e.g.][]{2010MNRAS.409L.104H,2017ApJ...850...29R}.

The dot-dashed line in Figure \ref{fig:radial}d shows the radial profile of the density-weighted average $\bar T_{\rm e}$  in our GRMHD solution for $R_\mathrm{high}=80$ and $R_\mathrm{low}=1$, which are typical values used in applications of the 
$R$-$\beta$ model. We see that at small $r$ the average $\bar T_{\rm e}$ reaches values several times higher than the level set by the energy balance of electrons in our ADAF solutions with the same values of $n$ and $B$. We note that the synchrotron cooling rate rapidly increases with temperature, approximately $\propto T_{\rm e}^5$, and therefore keeping electrons at such a high temperature (i.e., $\sim \bar T_{\rm e}$ found for the $R$-$\beta$ model) would require a heating rate orders of magnitude larger than the available accretion power. Furthermore, the $R$-$\beta$ model predicts significant emission at an extreme $\sim 10^{13}$ K, see Figure
\ref{fig:new}a. The problem of too high $T_{\rm e}$ in the $R$-$\beta$ model possibly occurs only in MAD solutions, where the plasma magnetization of the inner part of the flow is large, and $\beta$ is small.

The above issue is also apparent in figure 17 of \citet{2021ApJ...910L..13E}, comparing the temperature-ratio distribution from the radiative GRMHD simulations and the $R$-$\beta$ post-processing of the same GRMHD solution. In the innermost few $R_{\rm g}$ of the disk (which is the region where the production of radiation observed at millimeter and shorter wavelengths takes place in our GRMHD model), 
the radiative simulations (involving a detailed treatment of the thermodynamics of both ions and electrons) yield $T_{\rm e} \simeq 0.04 T_{\rm p}$ \citep[$\simeq 10^{11}$ K; see][for the same solution]{2019MNRAS.486.2873C}, which notably matches the temperature ratio given by the energy balance in our ADAF calculations. In contrast, the $R$-$\beta$ model gives an order of magnitude larger $T_{\rm e} \sim T_{\rm p}$ in the same region.

\subsection{SED modeling}

We now apply the GRMHD model to the SED observed from the nucleus of M87. 
We use the density, the electron temperature, and the components of the velocity and magnetic fields from the GRMHD simulation to compute the model spectrum, using the same GR Monte Carlo code as in the {\tt kerrflow} model. In it, we generate a large number of synchrotron photons according to the local values of plasma parameters and we track their motion in the Kerr metric, subject to Compton scattering competing with synchrotron self-absorption, until they cross the event horizon, escape to a distant observer, or get absorbed.

It is instructive to compare the formation of spectra in the GRMHD and ADAF models. 
The strongest radiation production occurs in the innermost part of the flow. It might therefore seem that the ADAF and GRMHD models which have similar values of $B$, $n$ and $T_{\rm e}$ in this region, also possess similar radiative properties. However, it turns out that the GRMHD model predicts a much weaker emission than ADAF. For example the GRMHD model A produces the synchrotron component almost by two orders of magnitude weaker than the ADAF models A1 and A9, despite having very similar parameters of the innermost emission region (see Figure \ref{fig:radial}). The reason for this is the much steeper $B$ profile in the GRMHD solution, which implies that the synchrotron radiation emissivity, $\propto B^2$, is much more concentrated near the event horizon and thus subjected to much stronger reduction by GR effects. In our GRMHD calculation, the {\it observed} radiation is reduced by over a factor of 10 (see Figure \ref{fig:sed}b and discussion below), as compared to a factor of $\sim 3$ in the ADAF model. Furthermore, the much faster decline of $B$ means that the amount of {\it produced} radiation is significantly smaller in the GRMHD model, because the  area of the disk where the synchrotron radiation is efficiently produced is smaller than in the ADAF model. Specifically, in our MHD models almost all radiation observed above the turnover frequency, $\nu_{\rm t}$ (below which the emission is strongly self-absorbed), is produced at $r \la 4$,  whereas in the ADAF models a significant contribution comes from the region extending out to $r \sim 20$. 

Another difference is related to the flow structure, which in the GRMHD solution is characterized by a range of parameter values at a given radius, related to the fluctuating nature of this solution. Such fluctuations are completely neglected in the analytical ADAF solutions, where at each $r$ the synchrotron emission is produced at a single value of $B$ and $T_{\rm e}$ and the resulting synchrotron spectrum has the shape of a relatively narrow bump peaking in IR, as seen in Figure \ref{fig:sed}a. Figure \ref{fig:new} shows the ranges of $T_{\rm e}$ and $B$ at which the synchrotron emission is effectively produced  (as estimated based on the local emissivity, $\propto T_{\rm e}^2 B^2 n$) in the GRMHD solution. We see, in particular, that a significant emission occurs at  $T_{\rm e}$ ranging from $\simeq 5 \times 10^{10}$ K to $\simeq 2 \times 10^{12}$ K.
Such a broad temperature spread gives rise to a broad synchrotron spectrum, extending from the millimeter to optical/UV band at an approximately constant level in $\nu F_{\nu}$ units,
as seen in Figures \ref{fig:sed}bc.

Figure \ref{fig:sed}b illustrates how the GR transfer of radiation affects the observed spectrum by comparing the spectra computed in model B neglecting (dashed line) and including (solid line) these transfer effects. As noted above, the synchrotron emission is strongly concentrated close to the BH and hence the observed spectrum is strongly reduced due to the bending of photon trajectories toward the BH and the capture of photons below the event horizon. Furthermore, a significant effect of the GR redshift can be seen, shifting the turnover frequency below the EHT frequency, whereas in the flat space-time it would be at $\simeq 400$ GHz. 

In model A the synchrotron component is weak and then it strongly underpredicts the observed SED, in particular at the EHT frequency.  The synchrotron emission can be enhanced by increasing the density, the electron temperature, and/or the magnetic field. However, a significant increase in the first two leads to an over-prediction of the X-ray data by the Compton component, whose amplitude relative to the synchrotron component increases with both density and temperature. The Thomson depth in the inner parts of the flow is small, $\sim 10^{-4}$, but in the Compton scattering at $\sim 10^{11}$ K the average energy increase is larger than a factor of $10^3$, and then the luminosity of the Compton component is comparable to that of the synchrotron. If the increase in either density or temperature is too large, the Compton component will be too strong. We then note that the model can be matched to the data, in the sense that the synchrotron spectrum agrees with the millimeter to UV data and the Compton spectrum does not exceed the X-ray data, only for a relatively strong $B$, corresponding to $b_{\rm mag} > 1$. These properties are illustrated in Figure \ref{fig:sed}c, which presents spectra in models C (dashed line) and D (solid line). Model C shows the combined effect of a large increase in $n$, to $\sim 10^6$ cm$^{-3}$ at $r \sim 2$, and an increase in $T_{\rm e}$ to $0.1 T_{\rm p}$. We see that whereas the synchrotron spectrum reproduces very well the lower energy data, the Compton emission exceeds the X-ray data by almost two orders of magnitude. Instead, the increase of $B$ in model D by $b_{\rm mag} \simeq 4$ (yielding the average $\beta \sim 1$ in the emitting region)
allows to fit the data between $\sim 10^{11}$ and $10^{15}$ Hz without exceeding the X-ray limit.

\begin{figure}
\centering
 \includegraphics[width=7cm]{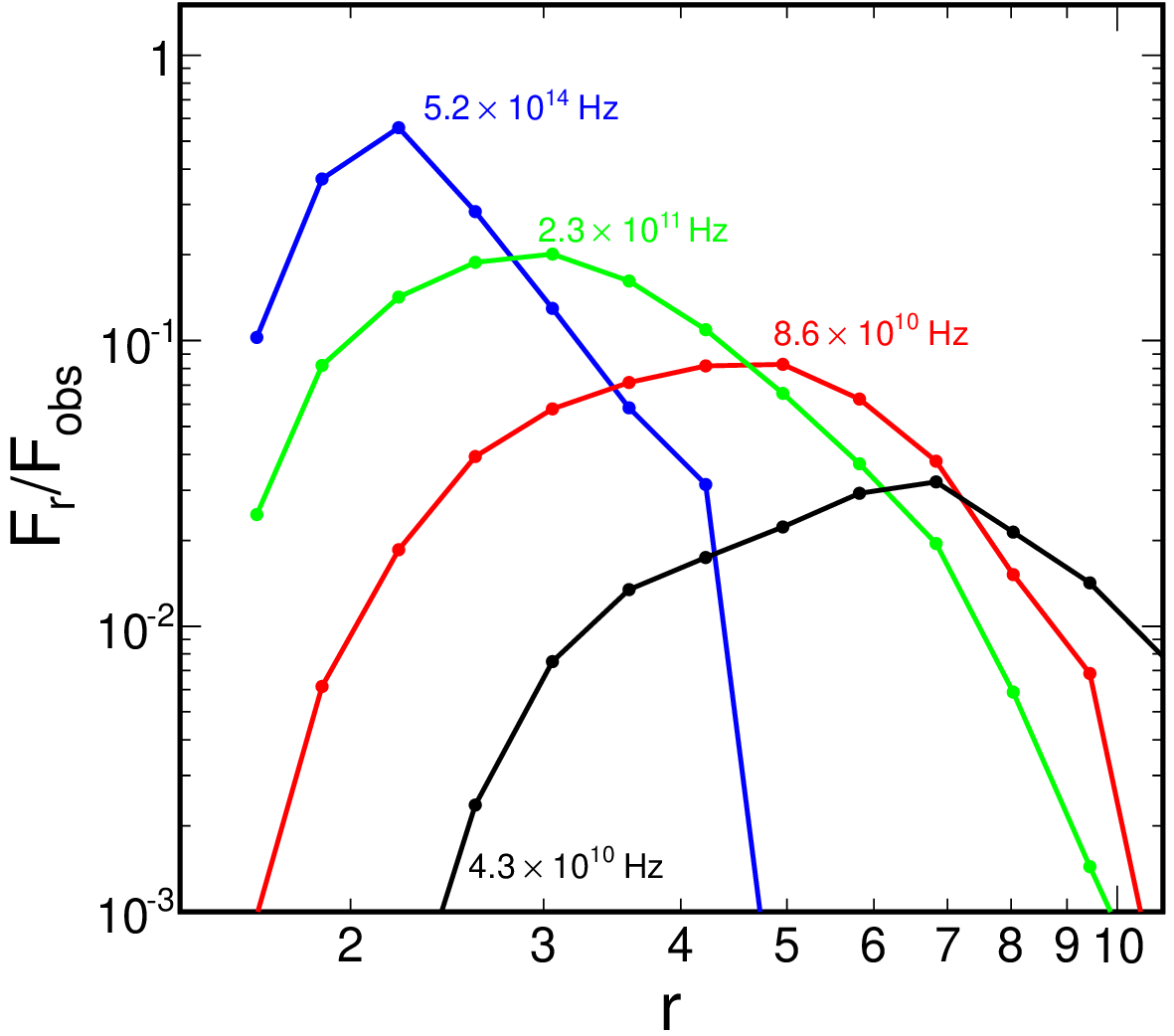}
\caption{The dots represent the contribution from a given radial bin (for logarithmic radial grid) to the flux observed at different frequencies at $i=17^{\circ}$ in model D, normalized by the flux measured by VLBA at $4.3 \times 10^{10}$ Hz, GMVA at $8.6 \times 10^{10}$ Hz, EHT at $2.3 \times 10^{11}$ Hz, and HST at $5.2 \times 10^{14}$ Hz, as reported by \citet{2021ApJ...911L..11E}.}
\label{fig:radial2}
\end{figure}

The bulk of radiation observed above $\nu_{\rm t} \simeq 2 \times 10^{11}$ Hz is produced in the disk region (i.e. near the equatorial plane) within $\simeq 4 R_{\rm g}$ (see Figure \ref{fig:radial2}). A small amount of radiation is produced beyond $4 R_{\rm g}$ due to the weakness of the magnetic field. Radiation produced within $2 R_{\rm g}$ is strongly reduced by GR transfer and we note that advection of photons into the BH is enhanced by the radial inflow of the radiating material. Therefore, at $\nu \gg \nu_{\rm t}$, the strongest contribution to the observed flux comes from $\simeq 2 R_{\rm g}$. At $\nu < \nu_{\rm t}$, the intrinsic absorption shifts the strongest contributing region to larger radii.

Finally, all the spectra discussed above were computed for the thermal distribution of electrons. In this version of the model, the high-energy component can be reduced below the observed X-ray flux, as in model D, but it cannot reproduce the X-ray spectrum. If the energy distribution of electrons is hybrid, the observed X-rays may be produced by non-thermal synchrotron emission. The dotted line in Figure \ref{fig:sed}c shows the synchrotron spectrum in model D assuming such a hybrid distribution of electrons, with the non-thermal component in the form of a high-energy power-law tail containing 5\% of energy of the total distribution. For the power-law index of the tail we assume $p = 3.2$, for which the photon index $\Gamma = 2.1$ is roughly consistent with that observed. We only note here an approximate agreement of such a model with the observed X-ray spectrum, which was derived in \citet{2021ApJ...911L..11E} assuming that the {\it XMM} and {\it NuSTAR} measurement can be represented by a single power-law. Formal spectral fitting of the X-ray data, performed in the detector space, is not feasible as our computational model is too time-consuming.

\section{Summary and discussion}

We compared radiative properties of two commonly used models of low-luminosity BH systems, the one based on the GRMHD simulation and the one using a semi-analytic ADAF description of the two-temperature accretion. The latter gives a similar distribution of the parameters of the flow to the distribution of the averaged parameters in the former, except for the magnetic field, which in the GRMHD solution has a much steeper radial profile corresponding to the amplification  of $B$ by the MRI, whereas the constant-$\beta$ structure typically assumed for ADAFs gives a much slower outward gradient of $B$. The rapid decline of $B$ with increasing $r$ in the GRMHD solution results in a much lower radiative efficiency, since it significantly limits the extent of the area where synchrotron radiation is produced and places this area in a region of strong reduction by GR effects. We note that the very steep profile is the typical property of GRMHD solutions with an initial configuration similar to ours, with a steep magnetic vector potential enabling the formation of the MAD state. However, this may be model dependent and relate to an unknown value of the $\beta$ parameter of the gas feeding the flow
\citep[see][]{2023MNRAS.521.4277R}.

Another difference between the two models is related to the neglect of the multidimensional structure of the flow in the ADAF description. This significantly affects the spectral shape, which in the ADAF model is represented by a relatively narrow synchrotron bump followed by similarly narrow scattering bumps. In contrast, the spread of parameters corresponding to the 3D GRMHD solution gives broad synchrotron as well as Compton components extending over many decades of photon energy.

We computed spectra for accretion rates and BH mass relevant for M 87*, also assuming a high value of the BH spin, $a = 0.9$, which may favor the production of the jet observed in this object. The ADAF model can explain the millimeter part of the spectrum observed in M87, but does not agree with the spectral data at other wavelengths. We note that previous ADAF models of this object are significantly affected by order-of-magnitude inaccuracies related to the neglect of the GR transfer of radiation as well as an oversimplified description of the Compton component.

The spectra based on the GRMHD solution are in a much better agreement with the observed SED. The low radiative efficiency of this model can be compensated by setting a relatively large $B$, corresponding to $\beta \sim 1$ at small $r$. An alternative way to increase the luminosity by increasing either the accretion rate (and hence the density) or the electron temperature results in a significant overprediction of the X-ray flux by the Compton component

The obvious caveat to our spectral modelling is that the SED data are obtained by observations with different angular resolution, which in the case of M87 translates to the spatial scale of $\sim 10 R_{\rm g}$ at the EHT frequency (i.e.\ 230 GHz) and $\ga 10^5 R_{\rm g}$ at higher frequencies, and therefore the optical, UV and X-ray radiation may be produced in a spatially distinct region from the emission region detected by  EHT. Nevertheless, the synchrotron component produced by predominantly thermal electrons in our GRMHD model is a good match for the spectral data between the millimeter and the UV range. 

The Compton component predicted in the GRMHD model is too hard to fit the data in the X-ray range. Nevertheless, comparison of this component with X-ray data imposes important constraints on the model and indicates that the emitting region must be characterized by a relatively large $\beta \sim 1$.
The observed X-ray emission may be produced within the inner few $R_{\rm g}$ by non-thermal synchrotron emission of a hybrid electron plasma with a high energy tail with the power-law index $\simeq 3.2$, which implies the acceleration index of $\simeq 2.2$ (taking into account the steepening of the electron distribution by synchrotron cooling).

The maximum contribution to radiation observed at 230 GHz in our best-matching solution (model D) comes from $\simeq 3 R_{\rm g}$, which region is characterised by the average $B \simeq 25$ G, $T_{\rm e} \simeq 6 \times 10^{10}$ K and $n \simeq 1.5 \times 10^5$ cm$^{-3}$. These values are consistent with parameters of the emission region estimated in \citet{2021ApJ...910L..13E} using a one-zone model, although we note that such one-zone estimations are significantly affected by the neglect of the GR transfer effects.
On the other hand, the accretion rate of our best-matching model, $\dot M \simeq 0.011 \, M_\odot$ year$^{-1}$, is larger than  $\dot M \la 0.002 \, M_\odot$ year$^{-1}$  assessed in the GRMHD models applied by \citet{2021ApJ...910L..13E} to explain the EHT observation. These models, however, assumed the $R-\beta$ prescription for the electron temperature, which strongly  affects the comparison with our results.

Our preferred value of $\dot M \simeq 0.01 \, M_\odot$ year$^{-1}$ is a factor of 10 lower than the Bondi accretion rate \citep{2003ApJ...582..133D}. This is sufficient to produce jet with a power on the order of $\sim 3\times 10^{43}$ erg$s^{-1}$, however exceeds the recent estimates by eg.\ \cite{2023A&A...677A.180P}.

At $\nu  \la 100$ GHz the synchrotron emission is strongly self-absorbed, which shifts the location of the effective flux maximum to larger radii, increasing with decreasing $\nu$  (Figure \ref{fig:radial2}). This radial shift is approximately consistent with the difference of sizes of the ring-like structures in the radio core of M87* resolved at 86 GHz and 230 GHz \citep{2023Natur.616..686L}.
Due to the strong self-absorption, the synchrotron emission from the inner accretion flow cannot explain the total flux measured at the low frequencies.
 This low-frequency radiation must then be generated, at least in part, on larger spatial scales, as indeed suggested by the core shifts observed in M87 at 86 GHz and lower frequencies \citep[e.g.][]{2021ApJ...911L..11E}. The contribution from the jet region is negligible in our calculations.
Some relevant physical processes, in particular internal shocks which can dissipate the bulk kinetic energy and channel some of it into particle acceleration \citep[e.g.][]{1978MNRAS.184P..61R,2001MNRAS.325.1559S}, are not handled accurately in current GRMHD codes, and therefore we have not attempted to model the jet emission that could explain the radio data. We also note that the dominance of emission from the disk region in our model is consistent, in particular, with the assessment 
of \citet{2023Natur.616..686L}
that the core emission in M87 comes from the accretion flow.

Our radiative calculations involve post-processing which is a rather crude approach, but is currently in use for practical reasons due to poor understanding of heating processes. The post-processing is not a well-established procedure.
We can expect the fluctuations seen in the MHD-flow structure to be reflected in the distribution of the electron temperature and the linear relation of the local $T_{\rm e}$ with the local $T_{\rm p}$, assumed in our post-processing, is the simplest way to account for this. Given a vague understanding of the physics of heating, the motivation to look for a more complex $T_{\rm e}/T_{\rm p}$ relation would be mainly observational, but we do not see it here, since our simple prescription gives a synchrotron spectrum that matches the observed SED.
Seemingly more sophisticated prescriptions are being used in this area of research, e.g.\ $R$-$\beta$, but using them without checking that the resulting $T_{\rm e}$ values are within physical limits, which is a common practice, can be very risky.
One way to check that $T_{\rm e}$ does not exceed the physically reasonable limits (which we intend to apply in our future works) is to use the general heating rate formulae \citep{2015MNRAS.454.1848R} and solve the electron energy balance assuming that a part of it goes to electrons; this would be analogous to the $\delta$-parametrization in the analytical ADAF models.

Interesting radiative GRMHD simulations of \citet{2019MNRAS.486.2873C} targeting M87, using specific prescriptions for electron heating based on either a Landau-damped turbulent cascade \citep{2010MNRAS.409L.104H} or magnetic reconnection \citep{2017ApJ...850...29R}, give similar to our model parameters in the inner disk, in particular, $T_{\rm e} \sim 10^{11}$ K. This, independently of the energy balance in the analytical model, motivated the choice of the scaling parameter (i.e., 0.04) in our $T_{\rm e}/T_{\rm p}$ relation. However, those simulations also give a much higher $T_{\rm e} \sim 10^{12}$ K in the strongly magnetized regions in the jet, and their spectra are fully dominated by these regions. Then, these spectra depend on an arbitrary cut in $\sigma$ more than on physical parameters \citep[as shown in figure 17 in][]{2019MNRAS.486.2873C}.

For our $T_{\rm e}$-prescription, the synchrotron radiation at $\nu \ga 10^{14}$ Hz is produced by electrons with $T_{\rm e} \sim 10^{12}$ K. The regions with such a large $T_{\rm e}$ represent a very small fraction of the total  flow volume, but their high emissivity makes them dominate the high-energy spectrum.
This high emissivity also implies a high cooling rate, which then requires enhanced heating to maintain the high temperature. 
Electrons in these high-$T$ regions may be more efficiently heated e.g.\ if the heating is due to magnetic reconnection, where electron heating is more efficient for a larger plasma magnetization, $\sigma \propto B^2/n$ \citep[e.g.][]{2017ApJ...850...29R}.
Indeed, the density in the high-$T$ regions in our solution is an order of magnitude lower, and $\sigma$ is similarly larger, than average in the disk.

\section*{Acknowledgements}
We thank the referee for useful comments that helped to improve the paper. We acknowledge support from the Polish National Science Center under grants 2023/48/Q/ST9/00138 and 2023/50/A/ST9/00527. We gratefully acknowledge Polish high-performance computing infrastructure PLGrid (HPC Center: ACK Cyfronet AGH) for providing computer facilities and support within computational grant no. PLG/2024/017013.

\bibliography{m87}{}
\bibliographystyle{aasjournal}
\end{document}